\documentclass[11pt]{article}
\usepackage[left=25mm, right=25mm, top=25mm, bottom=25mm, includehead=true, includefoot=true]{geometry}

\usepackage{graphicx}
\usepackage{url}
\usepackage{natbib} 
\usepackage{authblk} 
\usepackage[parfill]{parskip} 

\usepackage{sectsty}
\sectionfont{\normalsize}
\subsectionfont{\normalsize}
\subsubsectionfont{\normalsize}
\paragraphfont{\normalsize}

\usepackage[pdftex]{hyperref} 
\hypersetup{pdfborder={0 0 0} } 

\title{\vspace{-4cm}A scala library for spatial sensitivity analysis}

\author[1,2,3]{J. Raimbault\thanks{juste.raimbault@polytechnique.edu}}
\author[4,2]{J. Perret}
\author[2,3]{R. Reuillon}
\affil[1]{Center for Advanced Spatial Analysis, University College London}
\affil[2]{UPS CNRS 3611 ISC-PIF}
\affil[3]{UMR CNRS 8504 G{\'e}ographie-cit{\'e}s}
\affil[4]{LaSTIG STRUDEL, IGN, ENSG, Univ. Paris-Est}

\date{}

\begin{document}

\maketitle


\begin{abstract}
The sensitivity analysis and validation of simulation models require specific approaches in the case of spatial models. We describe the \texttt{spatialdata} scala library providing such tools, including synthetic generators for urban configurations at different scales, spatial networks, and spatial point processes. These can be used to parametrize geosimulation models on synthetic configurations, and evaluate the sensitivity of model outcomes to spatial configuration. The library also includes methods to perturb real data, and spatial statistics indicators, urban form indicators, and network indicators. It is embedded into the OpenMOLE platform for model exploration, fostering the application of such methods without technical constraints.

\end{abstract}


\section{Introduction}

The sensitivity of geographical analyses to the spatial structure of data is well known since the Modifiable Areal Unit Problem was put forward by \cite{openshaw1984modifiable}. This type of issue has been generalized to various aspects since, including temporal granularity \citep{cheng2014modifiable} or the geographical context more generally \citep{kwan2012uncertain}. When studying geosimulation models \citep{benenson2004geosimulation}, similar issues must be taken into account, extending classical sensitivity analysis methods \citep{saltelli2004sensitivity} to what can be understood as \emph{Spatial Sensitivity Analysis} as proposed by \cite{raimbault2019space}.

Several studies showed the importance of that approach. For example, in the case of Land-use Transport interaction models, \cite{thomas2018city} show how the delineation of the urban area can significantly impact simulation outcomes. \cite{banos2012network} studies the Schelling segregation model on networks, and shows that network structure strongly influences model behavior. The spatial resolution in raster configurations can also change results \citep{singh2007schelling}.

On the other hand, the use of spatial synthetic data generation is generally bound to model parametrization without a particular focus on sensitivity analysis, such as in microsimulation models \citep{smith2009improving}, spatialized social networks \citep{barrett2009generation}, or architecture \citep{penn2006synthetic}. \cite{raimbault2019space} however showed that systematically generating synthetic data, with constraints of proximity to real data configuration, can be a powerful tool to evaluate the sensitivity of geosimulation models to the spatial configuration.

This contribution describes an initiative to synthesize spatial sensitivity analysis techniques such as synthetic data generation, real data perturbation, and specific indicators, under a common operational framework. In practice, methods are implemented in the \texttt{spatialdata} scala library, allowing in particular its embedding into the OpenMOLE model exploration platform \citep{reuillon2013openmole}.

\section{Spatial sensitivity methods}

\paragraph{Generation of spatial synthetic data}

Realistic spatial synthetic configurations can be generated for geographical systems at different scales, and as different data types. Regarding raster data, (i) at the microscopic scale raster representation of building configurations (typical scale 500m) are generated using procedural modeling, kernel mixtures, or percolation processes \citep{doi:10.1162/isala00159}; and (ii) at the mesoscopic scale, population density grids (typical scale 50km) are generated using a reaction-diffusion urban morphogenesis model \citep{raimbault2018calibration} or kernel mixture. Regarding network data, synthetic generators for spatial networks include baseline generators (random planar network, tree network) and generators tailored to resemble road networks at a mesoscopic scale, following different heuristics including gravity potential breakdown, cost-benefits link construction, and a bio-inspired (slime mould) network generation model \citep{raimbault2018multi} \citep{raimbault2019urban}. Finally, regarding vector data, spatial fields generators can be applied at any scale (points distribution following a given probability distribution, or spatial Poisson point processes), while at the macroscopic scale system of cities with a spatialized network can be generated \citep{raimbault2020unveiling}.

\paragraph{Real data perturbation}

Real raster data can be loaded with the library and perturbed with random noise or following a Poisson point process. A raster generator at the microscopic scale can be used to load real building configurations from OpenStreetMap. For transportation networks, vector representations can be imported from shapefiles, directly from the OpenStreetMap API, or from a database (MongoDB and PostGIS are supported), and are transformed into a proper graph representation. Network perturbation algorithms include node or link deletion (for resilience studies e.g.) and noise on nodes coordinates.

\paragraph{Indicators}

Finally, various indicators are included in the library, which can be used to characterize generated or real configurations, and compare them. They include spatial statistics measures (spatial moments, Ripley K), urban morphology measures at the microscopic and mesoscopic scale, and network measures (basic measures, centralities, efficiency, components, cycles). Network measures can furthermore take into account congestion effects, as basic network loading algorithms (shortest paths and static user equilibrium) are implemented.

\paragraph{Implementation and integration in OpenMOLE}

The library is implemented in the language scala, which is based on the Java Virtual Machine and can benefit of existing Java libraries, and couples the robustness of functional programming with the flexibility of object-oriented programming. It can therefore easily be combined with one of the numerous Java simulation frameworks \citep{nikolai2009tools}, such as for example Repast Simphony for agent-based models \citep{north2013complex}, JAS-mine for microsimulation \citep{richiardi2017jas}, or Matsim for transportation \citep{horni2016multi}. The library is open source under a GNU GPL License and available at \url{https://github.com/openmole/spatialdata/}. A significant part of the library (synthetic raster generation methods) is integrated into the OpenMOLE model exploration platform \citep{reuillon2013openmole}. This platform is designed to allow seamless model validation and exploration, using workflows making the numerical experiments fully reproducible \citep{passerat2017reproducible}. It combines (i) model embedding in almost any language; (ii) transparent access to high performance computation infrastructures; and (iii) state-of-the-art methods for models validation (including design of experiments, genetic algorithms for calibration, novelty search, etc.). \cite{reuillon2019fostering} illustrates how this tool can be particularly suited to validate geosimulation models.

\section{Applications}

Different applications of the library have already been described in the literature. Regarding the generation of synthetic data in itself, \cite{doi:10.1162/isala00159} show that the building configuration generators are complementary to reproduce a large sample of existing configurations in European cities. \cite{raimbault2018calibration} shows that the reaction-diffusion morphogenesis model is flexible enough to capture most existing urban forms of population distributions across Europe also. \cite{raimbault2019second} shows that it is possible to weakly couple the population density generator with the gravity-breakdown network generator, and that correlations between urban form and network indicators can be modulated this way. \cite{raimbault2019urban} does a similar coupling in a dynamic way and shows that the co-evolution between road network and population distribution can be modeled this way.

For the application of the library to spatial sensitivity analysis, \cite{raimbault2019space} apply the population distribution generator to two textbook geosimulation models (Schelling and Sugarscape models), and show that model outcomes are affected by the spatial configuration not only quantitatively in a considerable way, but also qualitatively in terms of behavior of model phase diagram. \cite{raimbault2020unveiling} shows that the SimpopNet model introduced by \cite{schmitt2014modelisation} for the co-evolution of cities and transportation networks is highly sensitive both to initial population distribution across cities and to the initial transportation network structure.

\section{Discussion}


Beyond the direct application of the library to study the spatial sensitivity of geosimulation models, several developments can be considered. The inclusion of network and vector generation methods into OpenMOLE is currently explored, but remains not straightforward in particular because of the constraint to represent workflow prototypes as primary data structures, to ensure interoperability when embedding different models and languages. More detailed and operational transportation network capabilities are also currently being implemented into the library, including multi-modal transportation network computation and accessibility computation. Specific methods tailored for the validation of Land-use Transport Models are elaborated, such as correlated noise perturbation across different layers (coupling population and employment for example), or transportation infrastructure development scenarios. The strong coupling of generators into co-evolutive models such as done by \cite{raimbault2019urban} is being more thoroughly investigate in order to provide such coupled generators as primitives. This library and its integration with the OpenMOLE software should thus foster the development of more thorough geosimulation models validation practices, and therein strengthen the confidence in the results obtained with such models.


\end{document}